# The signature of orbital motion from the dayside of the planet τ Boötis b

Matteo Brogi[1], Ignas A. G. Snellen[1], Remco J. de Kok[2], Simon Albrecht[3], Jayne Birkby[1] and Ernst J. W. de Mooij[1,4]

[1] *Leiden Observatory, Leiden University, Postbus 9513, 2300RA Leiden, The Netherlands*
[2] *SRON, Sorbonnelaan 2, 3584 CA Utrecht, The Netherlands*
[3] *Department of Physics, and Kavli Institute for Astrophysics and Space Research, Massachusetts Institute of Technology, Cambridge, Massachusetts 02139, USA*
[4] *Department of Astronomy and Astrophysics, University of Toronto, 50 St. George Street, Toronto, ON M5S 3H4, Canada*

**The giant planet orbiting τ Boötis was among the first extrasolar planets to be discovered through the reflex motion of its host star[1]. It is one of the brightest known and most nearby planets with an orbital period of just a few days. Over the course of more than a decade, measurements of its orbital inclination have been announced[2] and refuted[3], and have subsequently remained elusive[4-8] until now. Here we report on the detection of carbon monoxide absorption in the thermal day-side spectrum of τ Boötis b. At a spectral resolution of R~100,000, we trace the change in the radial velocity of the planet over a large range in phase, determining an orbital inclination of $i$=44.5±1.5º and a true planet mass of 5.95±0.28 $M_{\rm Jup}$. This result extends atmospheric characterisation to non-transiting planets. The strong absorption signal points to an atmosphere with a temperature that is decreasing towards higher altitudes. This is a stark contrast to the temperature inversion invoked for other highly irradiated planets[9,10], and supports models in which the absorbing compounds believed to cause such atmospheric inversions are destroyed by the ultraviolet emission from the active host star[11].**

We observed τ Boötis for 3x6 hours during the nights of April 1, 8, and 14, 2011, with the CRyogenic InfraRed Echelle Spectrograph (CRIRES[12]) at the Nasmyth A focus of the Very Large Telescope UT1, located at the European Southern Observatory on Cerro Paranal, Chile. Targeting the planet almost continuously between orbital phase 0.37 < φ < 0.63, we collected 452 spectra at a resolution of R~100,000 in the wavelength range 2287-2345 nm, centred on the 2-0 R-branch of carbon monoxide. We used the ESO CRIRES data pipeline for the basic data reduction and the extraction of the one-dimensional spectra. We subsequently extracted the signal of the planet using purpose-built algorithms, similar to those utilised for high-dispersion transit spectroscopy of HD209458b[13], which are described in detail in the Supplementary Information (SI-2). The most critical step in this analysis is the removal of telluric features caused by the Earth's atmosphere, which completely dominate our spectra. This can be achieved without destroying the planetary signature only because the signal from the planet moves significantly in wavelength during our observations, due to its large change in radial velocity along the orbit.

We cross-correlated each of the 452 extracted and processed spectra with a CO template (described in SI-5.1), and subsequently aligned and combined all the cross-correlation functions, assuming a range of values for the maximum orbital radial velocity of the planet, $K_P$. Figure 1 shows the combined cross-correlation signal as function of $K_P$, corresponding to a wide range in orbital inclinations. We detect a 6σ absorption signal at $K_P = (110.0 ± 3.2)$ km sec$^{-1}$, at the systemic velocity of -16.4 km sec$^{-1}$, which is within the uncertainties of that of the host star[14] τ Boötis. The distributions of the values of the cross-correlated time series for points in the planet trail and out of the planet trail are shown in Figure 2. A detailed discussion on the noise properties of the cross-correlated time series and on the significance level of this detection is presented in SI-3. Figure 3 shows the radial velocity trail of the planet in CO over the full phase range of our observations. Combining the measured $K_P$ with the maximum radial velocity of the host star, $K_S = (0.4664 ± 0.0033)$ km sec$^{-1}$ (see SI-4), yields a star/planet mass ratio of 235.8 ± 7.1. With a host stellar mass[15] of $M_S = (1.34 ± 0.05)$ $M_{Sun}$ we derive a planetary mass of $M_P = (5.95 ± 0.28)$ $M_{Jup}$. In addition, using Kepler's Third Law, we determine a planet orbital inclination of $i = (44.5 ± 1.5)°$. Note that orbital solutions to the long-term monitoring of the radial velocity variations of the host star show a modest preference[16] for a slightly eccentric orbit with eccentricity $e = (0.023 ± 0.015)$. However, the combination of our planet radial velocity data and a reanalysis of the stellar data in the literature shows no evidence for an eccentric orbit (see SI-4). We therefore adopt a circular orbit for τ Boötis b in our analysis.

Spectro-polarimetric observations[14,17] of τ Boötis show that the host star exhibits strong differential rotation, with a period ranging from $P_{rot} = 3.0$ to 3.9 days from the equator to the poles. It indicates that the stellar rotation at intermediate latitudes is synchronised with the planet's orbital period ($P = 3.312$ days). If we assume that the stellar rotation axis is aligned with the normal to the orbital plane of the planet, the projected rotational velocity[16] of the star, $v \sin(i) \sim 15$ km sec$^{-1}$, combined with a stellar radius[15] of $R_S = (1.46 ± 0.05)$ $R_{Sun}$ and our measurement of $i$, indeed corresponds to a stellar rotational period of $P_{rot} = 3.3$ days, matching the planet's orbital period. Therefore, in addition to synchronization, it suggests that the orbital plane and the plane of stellar rotation are not significantly misaligned. We would like to point out that a large fraction of the hot Jupiters around hot stars ($T_{eff}$ > 6250 K) such as τ Boötis, whose orbital alignment can be measured via the Rossiter-McLaughlin effect, exhibit strong misalignments[18,19]. However, most of the massive planets ($M_p$ > 3 $M_{Jup}$) are found in more aligned orbits[20], and τ Boötis b does not break this trend.

Our observations at high spectral resolution are only sensitive to narrow spectral features, because of the particular data reduction necessary to remove the telluric contamination. Due to the high opacity at the wavelengths of molecular transitions, these narrow features probe the atmosphere at lower pressures than the surrounding continuum. The probed pressures are directly linked to the Volume Mixing Ratio (VMR) of CO, but the depth of the absorption features in the emitted planet spectrum depends on the relative temperatures at the levels of the continuum and CO lines. This means that there is a strong degeneracy between the temperature-pressure (T/p) profile of the planet atmosphere, and the VMR of CO. We compare our data with a range of models, in order to constrain the CO abundance and the T/p profile (see SI-5.2). We obtain a lower limit of the CO VMR by using the adiabatic lapse rate ($dT/d\log_{10}(p)$ ~ 1000 K at these temperatures), which is the maximum temperature gradient of a planet atmosphere before it becomes unstable to convection. An additional uncertainty is that the size of τ Boötis b is unknown, because it is a non-transiting planet. Since the average radius of the seventeen transiting hot-Jupiters currently known[21] with 3 $M_{Jup}$ < $M_P$ < 9 $M_{Jup}$ is 1.15 $R_{Jup}$, we assume this value for the planet. When we set the temperature of the atmospheric layer in which the continuum is formed to $T$ = 2000 K (near the expected dayside equilibrium temperature for a planet without energy redistribution to the night-side), and use an adiabatic lapse rate, we require a CO VMR of $10^{-5}$ to match the observed signal. If we assume a temperature of $T$ = 1650 K for the continuum photospheric layer (near the dayside equilibrium temperature for a planet with perfect redistribution to its night-side), under similar adiabatic conditions, a CO VMR of $10^{-4}$ is required. Note that this result is consistent with the metallicity of the host star, corresponding to a CO VMR of ~$10^{-3}$. We do not detect spectral features from methane or water vapour above a significance of 2σ, and we use our atmospheric models to derive upper limits to the relative abundances of these molecules of VMR($CH_4$)/VMR(CO) < 1 and VMR($H_2O$)/VMR(CO) < 5 at a 90% confidence level.

Photometric observations of hot Jupiters with the Spitzer Space Telescope have been interpreted as suggestive of thermal inversions, characterized by molecular features in emission rather than in absorption, of which HD209458b is the best-studied example. These inversions are likely fuelled by absorption of stellar radiation in a high-altitude absorbing layer. In such a model a thermal inversion is more likely to occur in the most highly irradiated planets, for which indeed some evidence exists[22]. The planet τ Boötis is more strongly irradiated than HD209458b. However, it is clear that τ Boötis does not exhibit a strong thermal inversion over the pressure range probed by our observations, since we see the CO signal in absorption. Although the exact pressure range probed depends on the CO abundance, the inversion layer invoked to explain the emission spectrum of HD209458b encompasses such a wide range in atmospheric pressures that it is evident that τ Boötis does not have a HD209458b-type thermal inversion. Interestingly, the host star of τ Boötis b exhibits a high level of chromospheric activity, and it has been recently suggested that hot-Jupiters orbiting active stars are less likely to have thermal inversions[11], because the strong UV radiation that accompanies chromospheric activity destroys the absorbing compound at high altitude, which would otherwise be responsible for the thermal inversion.

These observations show that high-resolution spectroscopy from the ground is a valuable tool for a detailed analysis of the temperature structure and molecular content of exoplanet atmospheres. The used technique not only reveals its potential for transmission spectroscopy[13], but also for dayside spectroscopy, meaning that atmospheric characterization is no longer constrained to transiting planets alone. Detection of different molecular bands will further constrain the relative molecular abundances and temperature-pressure profiles. In addition, tracing the signal along the orbit will reveal the planet phase function, which is linked to its global atmospheric circulation. Measuring this for different molecules may reveal changes between a planet's morning and evening spectrum driven by photo-chemical processes. Furthermore, molecular line profiles, in both dayside and transmission spectra, can potentially show the effects of a planet's rotational velocity, and unveil whether these hot Jupiters are indeed tidally locked.

**Acknowledgements** We thank the ESO support staff of the Paranal Observatory for their help during the observations. Based on observations collected at the European Southern Observatory (186.C-0289). S.A. acknowledges support by a Rubicon fellowship from the Netherlands Organisation for Scientific Research (NWO), and by NSF grant no. 1108595 "Spin-Orbit Alignment in Binary Stars".

**Author Contributions** M.B. led the observations and data analysis, and co-wrote the manuscript. I.S. conceived the project, contributed to the analysis and co-wrote the manuscript. R.d.K. constructed the planet atmosphere models. S.A. conducted the MCMC orbital analysis. J.B., E.d.M., R.d.K., and S.A. discussed the analyses, results, and commented on the manuscript.

**Author Information** The authors declare no competing financial interests. Correspondence and request for materials should be addressed to M.B. (e-mail: brogi@strw.leidenuniv.nl). Reprints and permissions information is available at www.nature.com/reprints.

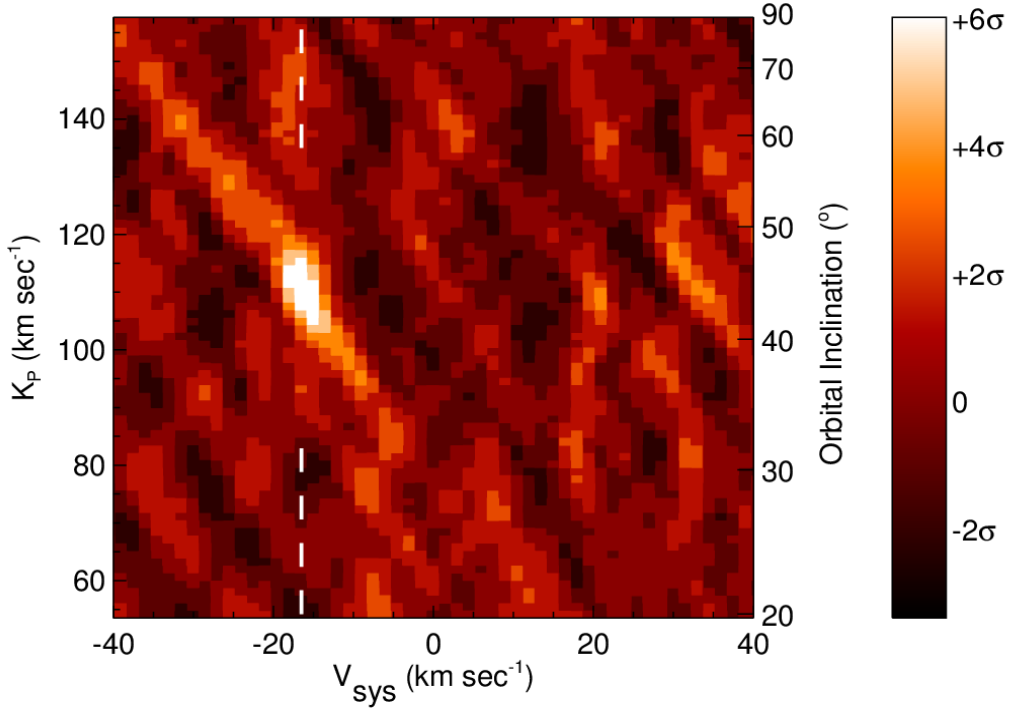

**Figure 1. CO signal in the dayside spectrum of the exoplanet τ Boötis b.** Colour scale plot of the carbon monoxide signal as function of heliocentric velocity on the *x*-axis, and the maximum radial velocity of the planet $K_P$ on the *y*-axis. The latter translates to an orbital inclination as indicated by the scale on the right side. Lighter colours indicate CO in absorption. A clear signal at a 6.2σ level is visible at the system velocity of τ Boötis (-16.4 km sec$^{-1}$), as indicated by the vertical dashed line, for a maximum planet velocity of $K_P$ = (110.0 ± 3.2) km sec$^{-1}$. This corresponds to an orbital inclination $i$ = (44.5 ± 1.5)° and to a planet mass of $M_P$ = (5.95 ± 0.28) $M_{Jup}$. The signal is obtained by cross-correlating a template spectrum of CO lines with the CRIRES/VLT spectra, which were each shifted in wavelength using the planet's ephemeris assuming a $K_P$. This to compensate for the changing Doppler effect caused by the change in the planet radial velocity over the large range in phase. The significance of the signal and the properties of the cross-correlated noise are discussed in the Supplementary Information.

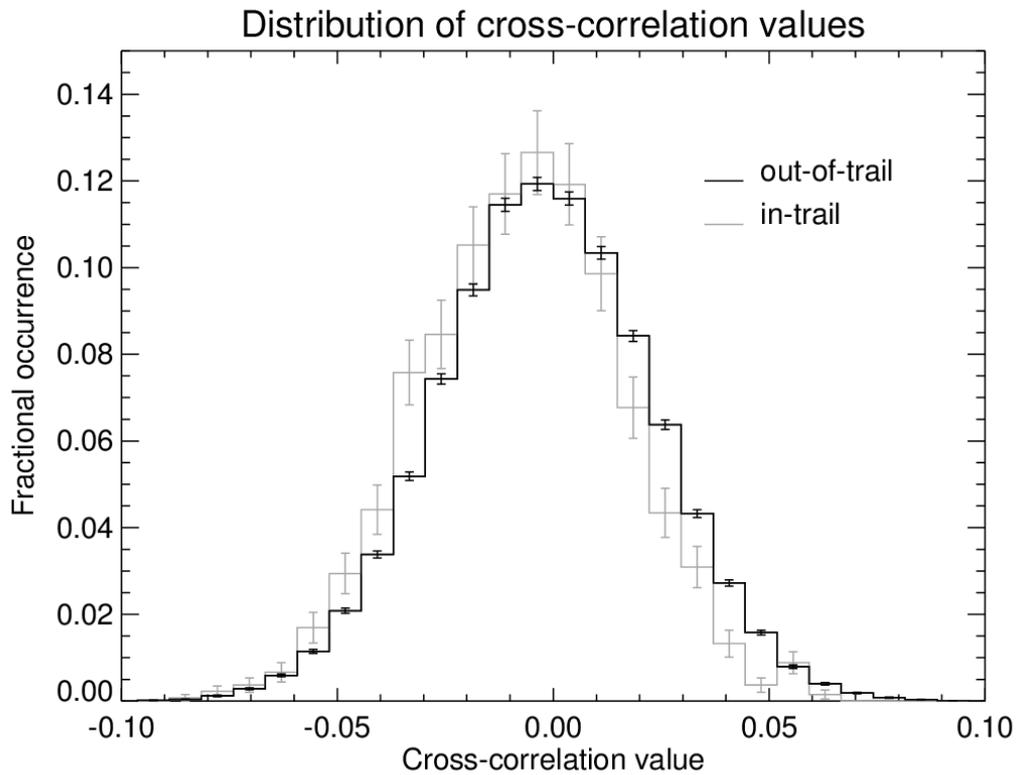

**Figure 2. Comparison of in-trail and out-of-trail cross-correlation values**. Distributions of the values of the cross-correlated time series for points in the planet trail (grey) and out of the trail (black). The error bars denote the square root of the number of data points in each bin (1σ). The two distributions clearly deviate, with the in-trail distribution shifted to lower pixel values due to the planet signal. A Welch t-test on the data rejects the hypothesis that the two distributions are drawn from the same parent distribution at the 6σ level.

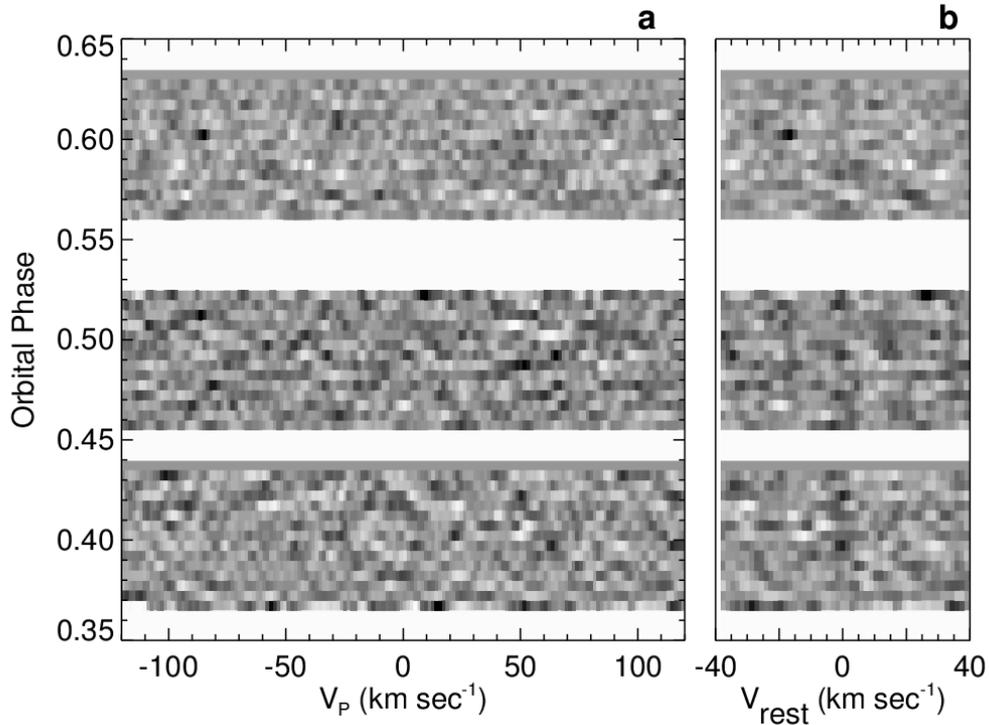

**Figure 3. The orbital trail of carbon monoxide absorption.** On panel **a**, the signature of CO in absorption is visible as a sinusoidal trace around the heliocentric radial velocity of τ Boötis from +80 km sec$^{-1}$ at phase 0.37, to -80 km sec$^{-1}$ at phase 0.63. On panel **b**, the data are shifted to the reference-frame of τ Boötis b, after subtracting the planet radial velocity computed assuming a circular orbit and a system inclination of 44.5°: here the planet signal is recovered as a vertical trace around $v_{rest}$ = 0 km sec$^{-1}$. A comparison between the observed trail and artificially generated data with the same noise properties is shown in the Supplementary Information (Fig. S4).

# SUPPLEMENTARY INFORMATION

## SI-1. The CRIRES observations

We observed τ Boötis during the nights of April 1, 8 and 14, 2011, using the CRyogenic high-resolution Infrared Echelle Spectrograph (CRIRES)[12], located at the Nasmyth A focus of the ESO Very Large Telescope UT1 (Antu), in conjunction with the Multi-Application Curvature Adaptive Optic system (MACAO)[23]. We collected 452 spectra over ~18 hours of observations, covering most of the planetary orbit from phase φ = 0.37 to φ = 0.63 (Fig. S1). CRIRES utilises four 1024x512 pixel Aladdin III detectors, with a gap of ~250 pixels between each chip. Our observations covered a spectral range of 2287.5 nm to 2345.4 nm, targeting the 2-0 R-branch of carbon monoxide, with a 0.2" slit, resulting in a resolution of R ~ 100,000 per frame. Accurate background subtraction requires nodding of the telescope over 10" between frames along the direction of the slit. Due to the uncorrected distortion of the spectrograph, the combination of the spectra at the two nodding positions reduces the resolution to R ~ 87,000.

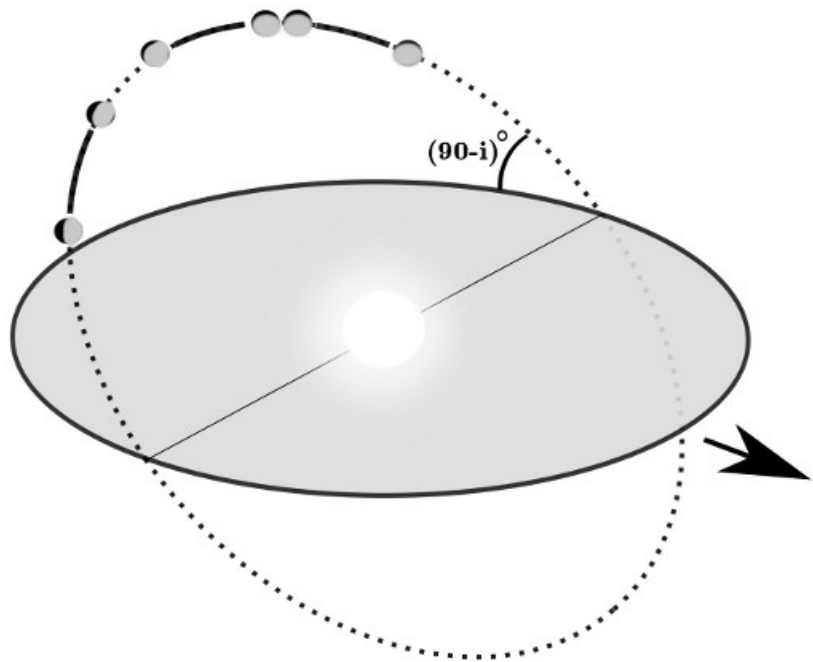

*Fig. S1. Schematic representation of the planetary system τ Boötis.* *We obtained observations of τ Boötis during three nights in April 2011, covering a large range in orbital phase, as indicated by those parts of the orbit drawn with a thick line and with planet-images marking the start and endpoints. The arrow indicates the direction of the Earth, from which we see the orbit at an inclination i. The planet orbit and host star are drawn approximately at the correct scale, while the planet-images are enlarged by a factor ~3 for clarity.*

## SI-2. Data reduction and analysis

**SI-2.1 - Initial data reduction**

For the data reduction and analysis, we followed a similar procedure as used for the detection of carbon monoxide in the transmission spectrum[13] of HD 209458 b. As a first step, we utilised the CRIRES pipeline v. 2.1.1 for the dark subtraction, flat-fielding, non-linearity and bad-pixel corrections, and the extraction of the one-dimensional spectra. For the remainder of the analysis, we developed purpose-built algorithms using the Interactive Data Language (IDL). We first corrected for additional hot pixels not picked up by the pipeline, and masked bad regions in the arrays. Next, for each night we chose the spectrum with the highest signal-to-noise ratio as a reference, and aligned all the other spectra to it, by fitting the positions of a set of strong lines in each spectrum and applying a global solution through spline interpolation. In this way, we achieved a typical RMS of ~0.1-0.2 pixels per spectrum in the residual line position, with respect to a common wavelength scale. We subsequently determined the wavelength solution by comparing the positions of the telluric lines with those in the HITRAN[24] database. The typical uncertainty in the absolute wavelength solution is estimated to be 0.15-0.20 km sec$^{-1}$. We finally normalised each spectrum to the continuum level of the reference, by fitting a second order polynomial to the flux ratios as function of wavelength, being careful not to sample regions near the telluric absorption features.

**SI-2.2 - Removal of the telluric line contamination**

After the initial data reduction, all the spectra have the same continuum level, while the depths of the telluric lines vary by 15-20%. This is primarily caused by changing airmass, but also due to variations in seeing, instrument resolution, in-slit guiding and adaptive optics corrections, which influence the shape and depth of the absorption lines. For each of the four arrays, the spectral series are handled and visualised as a two-dimensional matrix (Fig. S2, panel *a*), in which the horizontal axis is pixel position (representing the wavelength) and the vertical axis is frame number (representing the orbital phase or time). The data are dominated by the absorption lines formed in our own atmosphere, which are fixed in wavelength, and consequently fall on fixed columns in the matrix. Conversely, any signal from the planet will be Doppler-shifted due to the changing radial component of its orbital velocity, and its spectral signature will appear as a tilted trace across the matrix[25]. This is a key-point in the data analysis, allowing the planet and the telluric signal to be disentangled.

We remove the telluric contamination as follows. The fluxes of each column of the matrix are first fit as a linear function of the geometric airmass at the time of observation, which reduces the variations per column to a few per cent or less (Fig S2, panel *b*). The remaining time-correlated residuals in the matrix are mostly due to the varying conditions in the Earth's atmosphere throughout the night. We measured the flux of few of the deepest $H_2O$ and $CH_4$ lines over time, and used these to correct the rest of the matrix with a column-by-column linear regression. To remove the remaining low-order structure across the matrix, we also applied a high-pass filter to each row. After these steps, each point in the matrix is normalised by its uncertainty (Fig S2, panel *c*) so that noisy parts of the data do not dominate the cross-correlation. By removing the time-dependence of the signal in each column of the matrix, we get rid of the telluric contamination, while the planetary signature is left largely unaffected. Note that, since τ Boötis is a F7V star, it does not exhibit significant stellar spectral features in the observed wavelength range.

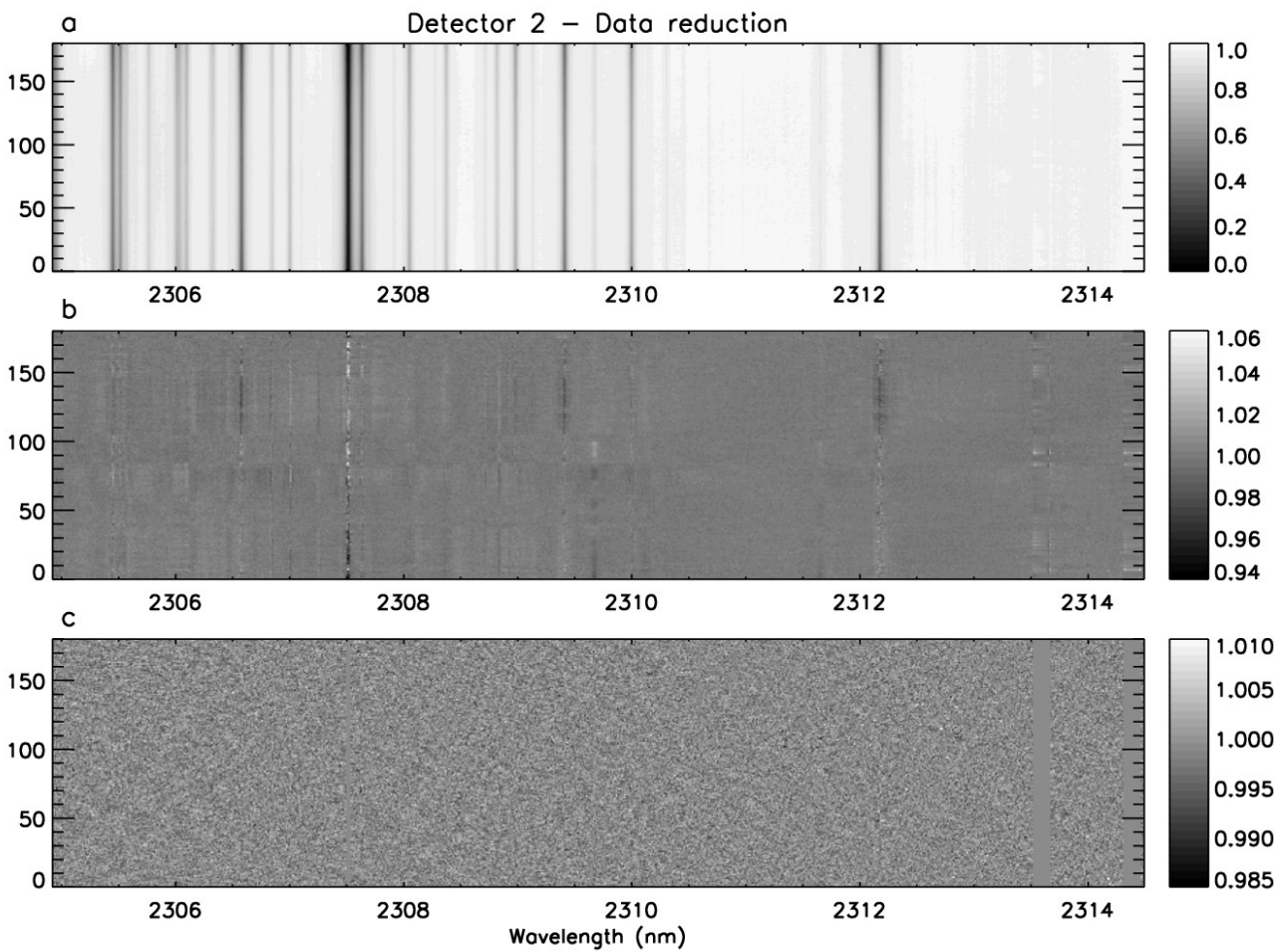

*Fig. S2: An example of our data reduction chain, showing the spectral series from CRIRES detector #2, taken on April 1, 2011. The y-axis corresponds to time. Data are first aligned and normalized to the same continuum level (a), and subsequently de-trended from the effects of the geometric airmass (b). Finally, correlated and low-order residuals are removed and bad regions in the array are masked (c). See section SI-2.2 for further details.*

## SI-2.3 - Cross correlation and signal extraction

As a result of the data processing described above, the data are no longer sensitive to any broad spectral features. However, they are still sensitive to narrow absorption or emission lines in the planet spectra – in this case, tens of CO lines. In order to combine the signal from all the CO lines, we cross-correlate each spectrum with a CO template, which is shifted in velocity from -150 km sec$^{-1}$ to +150 km sec$^{-1}$ (in steps of 1.5 km sec$^{-1}$), covering all the possible radial velocities for the planet. The template is extracted from a model spectrum for the atmosphere of τ Boötis b (see Section SI-4.1 for details), by fitting the positions and amplitudes of the strongest lines, and redefining them as a series of narrow Gaussian profiles. In this way the broadband component of the model spectrum is discarded, and an optimal sampling for the signal retrieval is achieved. Note that, since the expected carbon monoxide signal in the wavelength regions covered by detectors #1 and #4 is only few per cent of the total, and that in addition these two detectors are significantly affected by an uncorrected odd-even effect[26], we only include detectors #2 and #3 in our analysis. The result of the cross-correlation across the three nights for detectors #2 and #3 combined is shown by the matrix in Figure S3, where the horizontal axis represents radial velocity while the vertical axis is now the orbital phase. When the Doppler-shifted model matches the data, a positive (or negative) value of the cross-correlation will denote emission (or absorption) from the atmosphere of the planet. With about 25-30 CO lines falling in the observed wavelength range, the signal resulting from the cross-correlation is at least a factor of 5 stronger compared to that of a single CO line, but is still buried in the noise at this stage.

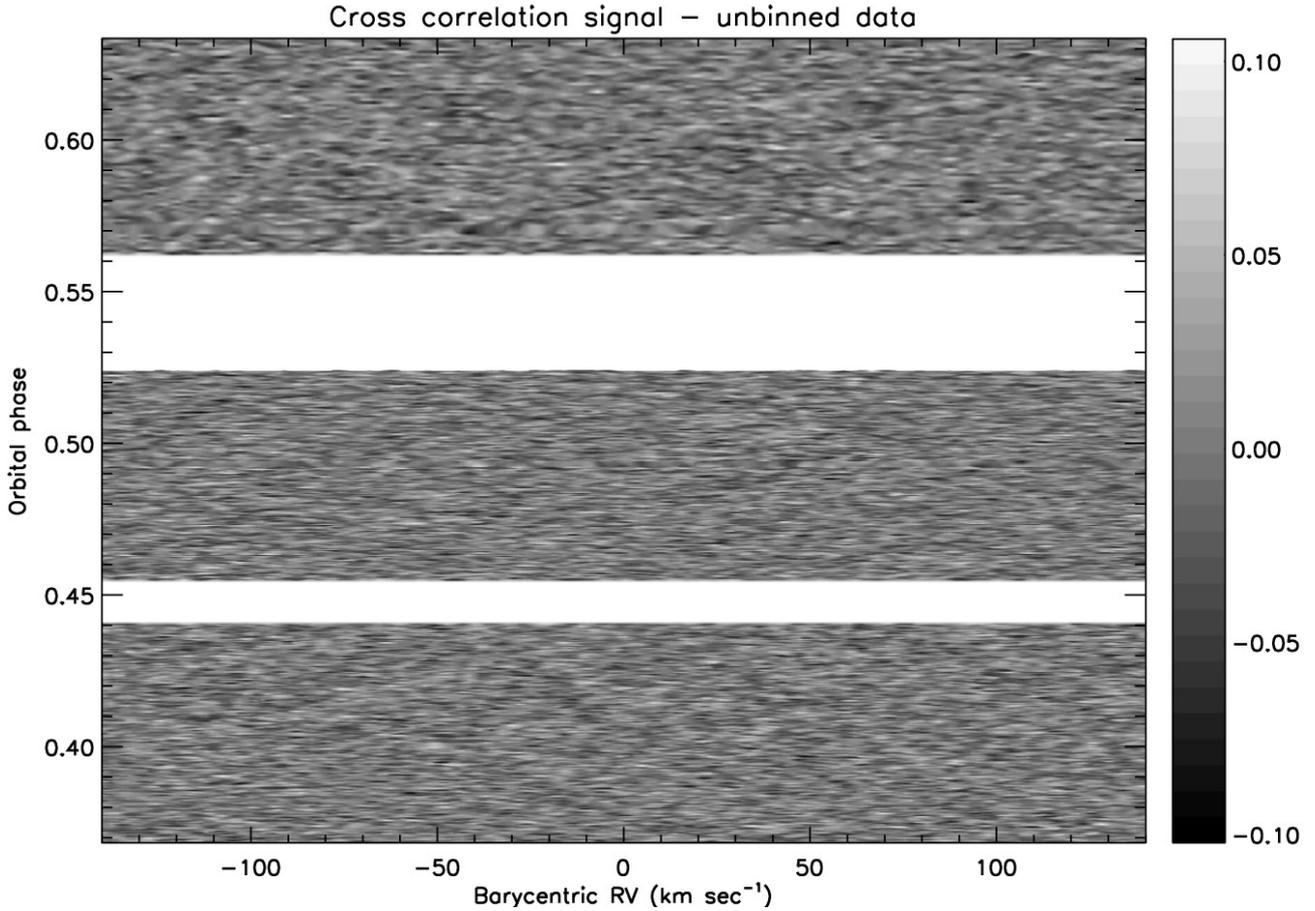

*Fig. S3: Result of the cross-correlation between our data and a template for carbon monoxide, as a function of barycentric radial velocity and planet orbital phase. In this way, the signal from 25-30 molecular lines is combined. The planet signal is not visible, and only becomes evident after binning the data in time (see Fig. 3 of the main paper). Note that the grey-scale represents the cross-correlation values in arbitrary linear units.*

From previously measured orbital parameters[16] of τ Boötis b, we determine the phase of the planet at the time of our observation with an uncertainty of ±0.012, while the orbital radial velocity of the planet $K_P$ is dependent on the unknown system inclination. We therefore assume a large range of values for $K_P$ and a small global phase shift $\Delta\varphi$ (see SI-3), and determine the expected radial velocity curve for each one of them, incorporating the heliocentric velocity of the observer (due to the orbital and rotational motion of the Earth). Subsequently, we recompute the cross-correlation matrix in the rest frame of the planet by a linear interpolation of each spectrum, and sum along the vertical (time) axis, thus increasing the signal in the cross-correlation function by a factor of $N^{1/2} \sim 20$, where $N$ is the total number of frames. This sum is performed independently for each array of the CRIRES detector, and each night of observation. The signals from the individual arrays are then combined with a weighting determined from the CO template, by summing in quadrature the depths of the model lines falling in the wavelength ranges of the two detectors. Finally, the combined cross-correlation signal of the three nights is summed.

# SI-3. The CO detection and the planet velocity trail

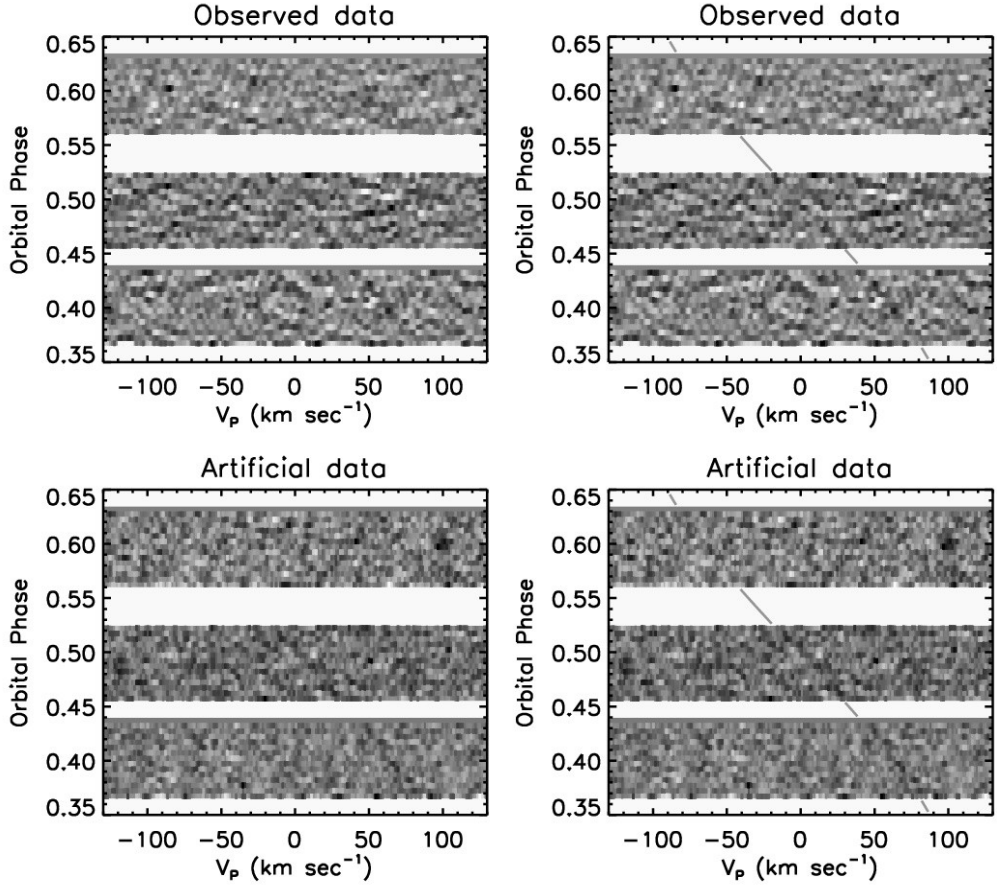

*Fig. S4, top panels: the observed velocity trail of τ Boötis b binned to 0.005 in phase, with and without the guiding lines showing the planet radial velocity (right and left panel respectively). Bottom panels: artificial data consisting of the injected planet velocity trail plus Gaussian noise at the same level as in the observed data. It shows that the planet trail is visible in the observed data at the expected level.*

In Figure 3 of the main paper we show the resulting carbon monoxide velocity trail in the cross-correlation data, binned to 0.005 in phase. In Figure S4 the same data is shown with and without the grey line guiding the eye (top right and left panel). This is compared with a signal of the same strength, embedded in artificial data with the same noise properties (bottom right and left panel). For a signal at a 6σ level, integrated over the entire time series, it is expected that on average at each binned point in phase the velocity trail is visible at a signal-to-noise of ~1. The comparison with the artificial data indicates that the velocity trail is visible at this expected level. Although it is not essential to visually see the trail at this intermediate stage, it is an extra safety check against a spurious signal originating from strong residuals in one or a few spectra in the dataset.

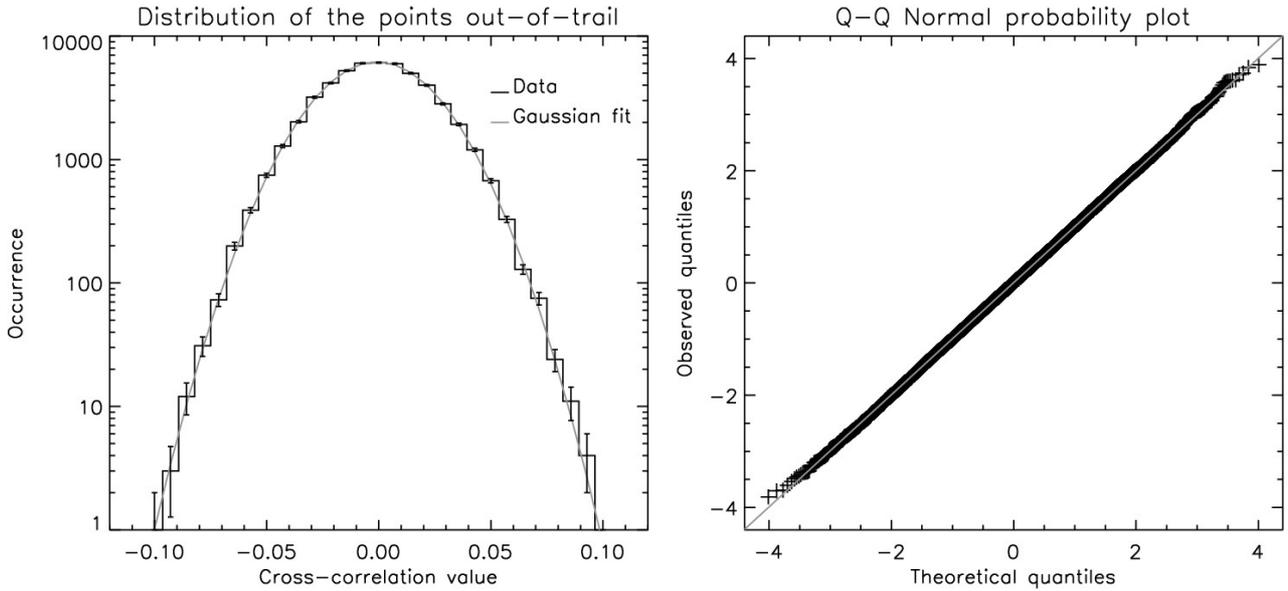

*Fig. S5, left panel: Distribution of the values of the cross-correlated time series, after excluding the points belonging to the planet trail. A Gaussian curve with zero mean is fitted to the data and overplotted in light grey. The two distributions agree across the entire range that can be tested given the total number of points. The latter corresponds to ~4 times the standard deviation of the fitted Gaussian. Right panel: The Q-Q normal probability plot[27], showing the quantiles of the cross-correlation values against the quantiles of a normal distribution. The tight correlation indicates no sign of non-Gaussianity in the data.*

By comparing the peak value of the total cross-correlation signal with the standard deviation of the noise (Fig. 1 of the main paper), we determine the signal-to-noise ratio of the CO absorption to be 6.2. In the left panel of Figure S5 we show the noise distribution of the cross-correlation time-series (i.e. the matrix in Figure S3). Overplotted is the best-fit Gaussian distribution with a mean pixel value fixed to zero. The data is well fitted by the Gaussian distribution down to ~4 times the standard deviation. There are not enough data points to test Gaussianity beyond this limit. We also compared the cumulative density function of the observed cross-correlation values to that of a normal distribution using a Q-Q probability plot[27], shown in the right panel of Figure S5. In such a plot, the quantiles of the observed distribution are plotted against the expected values for a normal distribution, allowing a visual diagnostic of the matching between the two. In our case, the two sets of quantiles are well correlated across the entire range, meaning that the two cumulative density functions agree between -4σ and +4σ. Note that it is not surprising that the distribution of the cross-correlation values is Gaussian, because although the observed spectra may have varying noise patterns as function of wavelength (due to the telluric absorption), each pixel value is first normalized by its uncertainty. In addition, the cross-correlation with a template, consisting of tens of CO lines spanning the entire wavelength range, also strongly scales down any systematic which may deviate from a Gaussian distribution.

After verifying that our data have Gaussian noise properties, we tested the statistical significance of the CO absorption. In Figure 2 of the main paper, we show the distribution of pixel values inside and outside the velocity trail of τ Boötis b. It suggests that the dispersions in the two distributions are similar, but that the points inside the velocity trail are shifted to lower values. We quantified this using the Welch t-test[28], which assumes that both groups of data are sampled from Gaussian populations (like in a standard t-test), but which does not requires that they have the same sample variance. The hypothesis that the two distributions are drawn from the same parent distribution is rejected at the 6σ confidence level, in line with our earlier estimate.

As an additional test, we removed the planet signal using the best-fitting model and the planet orbital solution (as explained in section SI-5) and re-injected it at random systemic velocities between -45 and +45 km sec$^{-1}$, and $K_P$ in the range 100–120 km sec$^{-1}$. In this way we sampled different regions of the spectral series and constructed the distribution of the retrieved signal, shown in Figure S6. This results in a mean signal-to-noise of 6.3 with a standard deviation of 1.0, meaning that the measured absorption signal is again consistent with a signal-to-noise of ~6, regardless to where the planet signal is in the data.

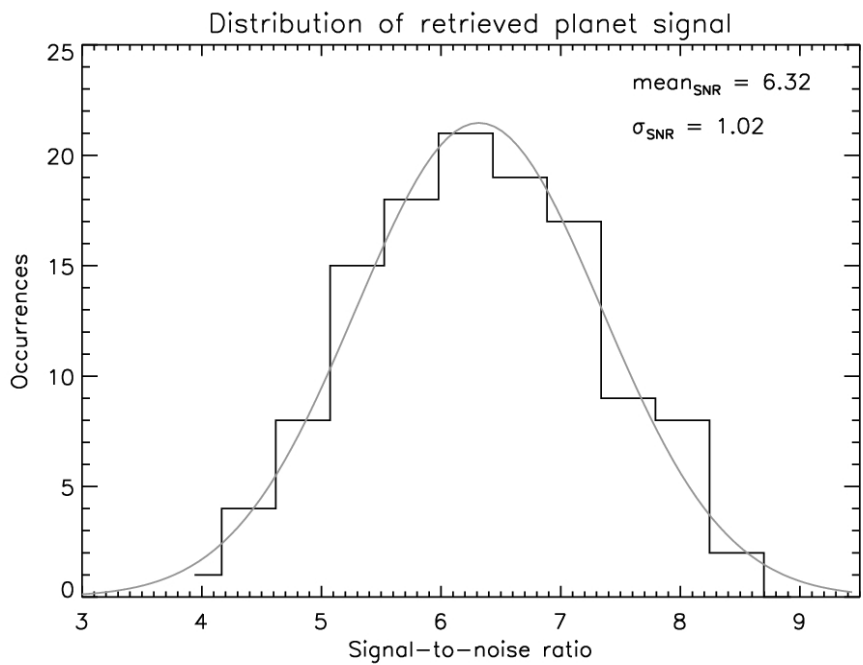

*Fig. S6: Distribution of the retrieved signal-to-noise ratios of artificially injected planet signals with strengths equal to the observed one, but with a range of systemic and planet orbital velocities. The mean (6.3) and the standard deviation (1.0) are those expected for a signal-to-noise ratio of ~6, meaning that ~66% of the occurrences are between 5.3σ and 7.3σ, independent from where the planet signal is injected.*

We shifted and co-added the 452 reduced spectra of τ Boötis, according to the planet ephemeris and inclination as derived below, to construct a 1D-spectrum. This is the upper spectrum shown in Figure S7, for both arrays separately. For comparison, the best-fit model spectrum of the planet atmosphere is shown in the middle, and the same model spectrum with artificial Gaussian noise added at the measured level is provided below. Since the total cross-correlation signal is measured at a 6σ level, the strongest lines should be visible at ~1σ, but also many features at the same level should not correspond to genuine absorption lines, which is indeed what is seen in both the real and artificial data. Again, this shows that the cross-correlation signal does not originate from one particular part of the spectrum, but from the summed contribution of all absorption lines.

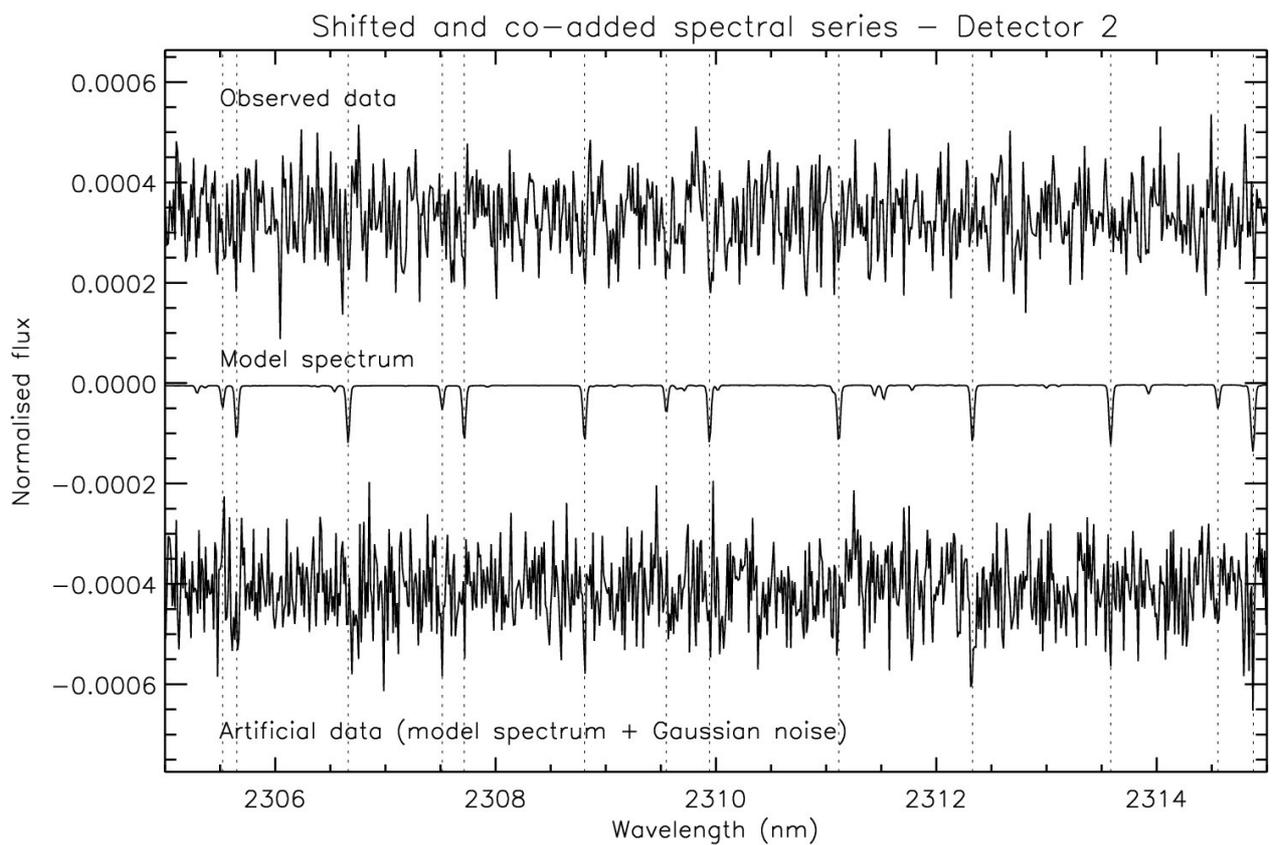
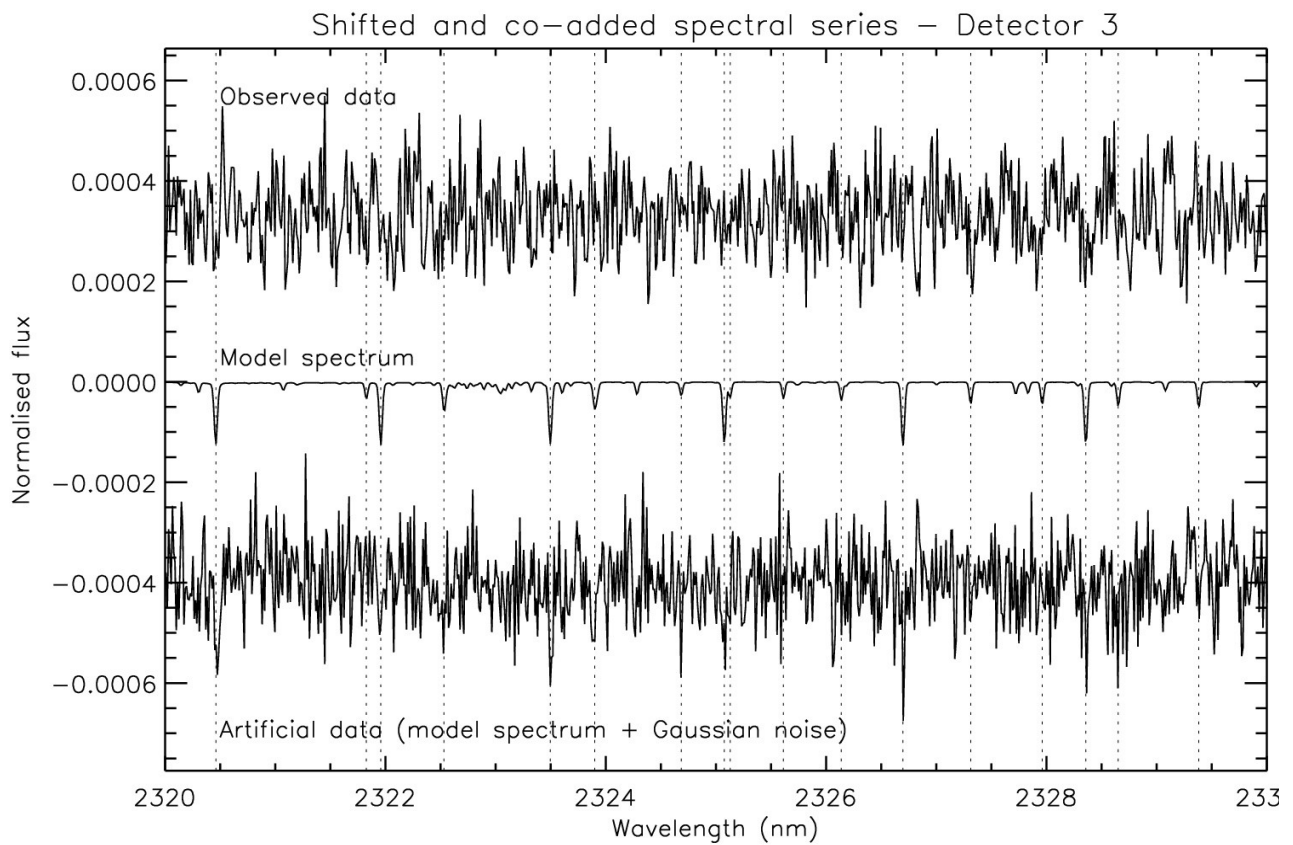

*Fig. S7: One-dimensional spectrum of τ Boötis b, constructed by shifting and adding the individual spectra according to the planet orbital solution, with the top and bottom panels showing data from detectors 2 and 3 respectively. In both panels, the model spectrum and model spectrum + Gaussian noise are shown for comparison. The dotted vertical lines indicate the positions of the strongest CO lines.*

# SI-4. The planet orbit and mass

The orbital solution for τ Boötis b presented in the literature, as derived from the radial velocity variations of the host star[16], shows a weak preference for an eccentric solution, with eccentricity $e = (0.023 \pm 0.015)$ and longitude of periastron $\omega \approx 188°$. Adopting this solution, the signature of carbon monoxide from the atmosphere of the planet is retrieved in absorption for a maximum planet radial velocity of $K_P = (110.2 \pm 3.2)$ km sec$^{-1}$ and at a 6.2σ confidence level. The significance of the planet signal is not affected by the value of Δφ across a small range. Therefore, this parameter can be arbitrarily set, as long as it is consistent with the uncertainties in the orbital solution. With a value of Δφ = 0.0068, the position of the CO signal matches the systemic velocity of τ Boötis, $V_{sys} = (-16.4 \pm 0.1)$ km sec$^{-1}$, determined by extrapolating a recent literature value[14] with our measured linear trend of $\Delta V_{sys} = (0.0458 \pm 0.0033)$ m sec$^{-1}$ day$^{-1}$ (see below).

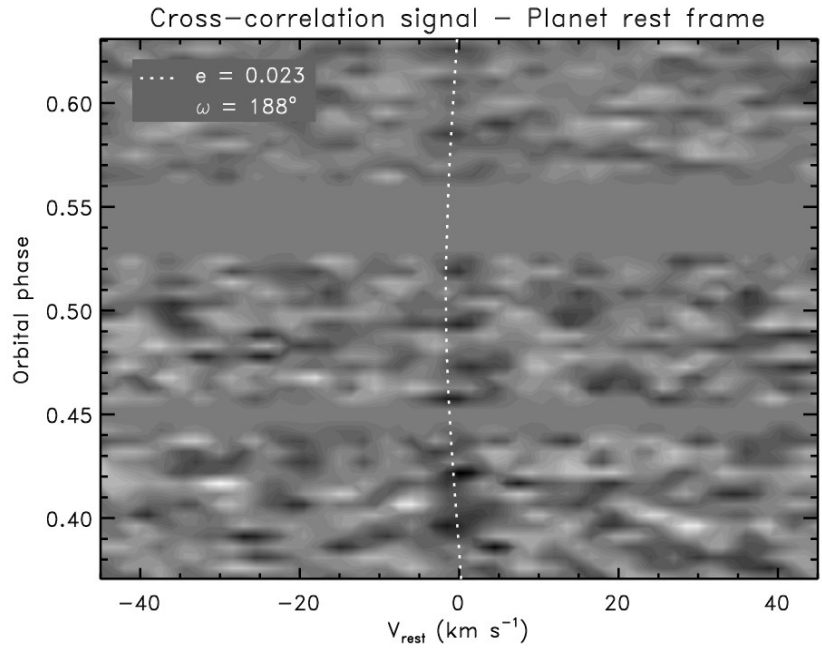

*Fig. S8: Radial velocity trail of τ Boötis b, shown in the rest frame of the planet assuming a circular orbit, and binned in phase to increase the signal-to-noise ratio. The eccentric (e = 0.023, ω = 188°) solution is shown by the dotted white curve for comparison.*

After performing the analysis assuming an elliptical orbit for the planet, we checked whether our data support an eccentric solution. Fig. S8 shows the observed radial velocity trail of τ Boötis b in its rest frame, assuming a circular orbit. The dotted line indicates the expected planet radial velocity for the eccentric solution (e = 0.023, ω = 188°). Although not statistically significant, at least by eye it seems that the eccentric solution, which is slightly curved with respect to the circular solution, is a poorer match to the planet radial velocity trail. Therefore, we independently re-analysed all of the available radial velocity data of the host star, obtaining solutions for the time of inferior conjunction ($T_0$), the orbital period ($P$), the stellar radial velocity semi-amplitude ($K_S$) and the residual linear trend on the data ($\Delta V_{sys}$), which were found to be all consistent with the literature values[16]. However, we find that the early data taken between 1987 and 1994, which are significantly noisier than the more recent measurements, are the cause of the slightly eccentric solution. The left panel of Fig. S9 shows $e$ versus $T_0$ for our MCMC analysis using all the data, showing the weak preference for an eccentric solution and a correlation between $e$ and $T_0$. The right panel shows the same MCMC analysis when excluding the data prior to 1995. No correlation between $e$ and $T_0$ is seen and the preference for an eccentric solution is no longer present. Due to the combination of the planet radial velocity data and our re-analysis of the available stellar radial velocity data, we opt to use the circular orbital solution. In this way, a 6.2σ absorption signal from carbon monoxide is observed at the maximum planet radial velocity of $K_P = (110.0 \pm 3.2)$, as shown in Figure 2 of the main paper, suggesting that the impact of the circular solution on our analysis is negligible. The signal is detected at the systemic velocity of τ Boötis for a phase shift of $\Delta\varphi = -0.0091$, again consistent within the 1σ uncertainty. Combining the stellar and planet radial velocity, yields a new set of orbital parameters with $P = 3.312433(19)$ days, $T_0 = (2,455,652.108 \pm 0.004)$ HJD, $K_S = (0.4664 \pm 0.0033)$ km sec$^{-1}$. We determine the star/planet mass ratio to be $K_P/K_S = (235.8 \pm 7.1)$, which together with the estimated[15] stellar mass of $M_S = (1.34 \pm 0.05)\ M_{Sun}$ gives a planet mass of $M_P = (5.95 \pm 0.28)\ M_{Jup}$. Knowing the mass ratio and the orbital period of the planet, we apply Kepler's Third Law to derive a planet orbital velocity of $V_P = (157.0 \pm 2.6)$ km sec$^{-1}$. The ratio $K_P/V_P = (0.701 \pm 0.024)$ is the sine of the orbital inclination, which we determine to be $i = (44.5 \pm 1.5)°$.

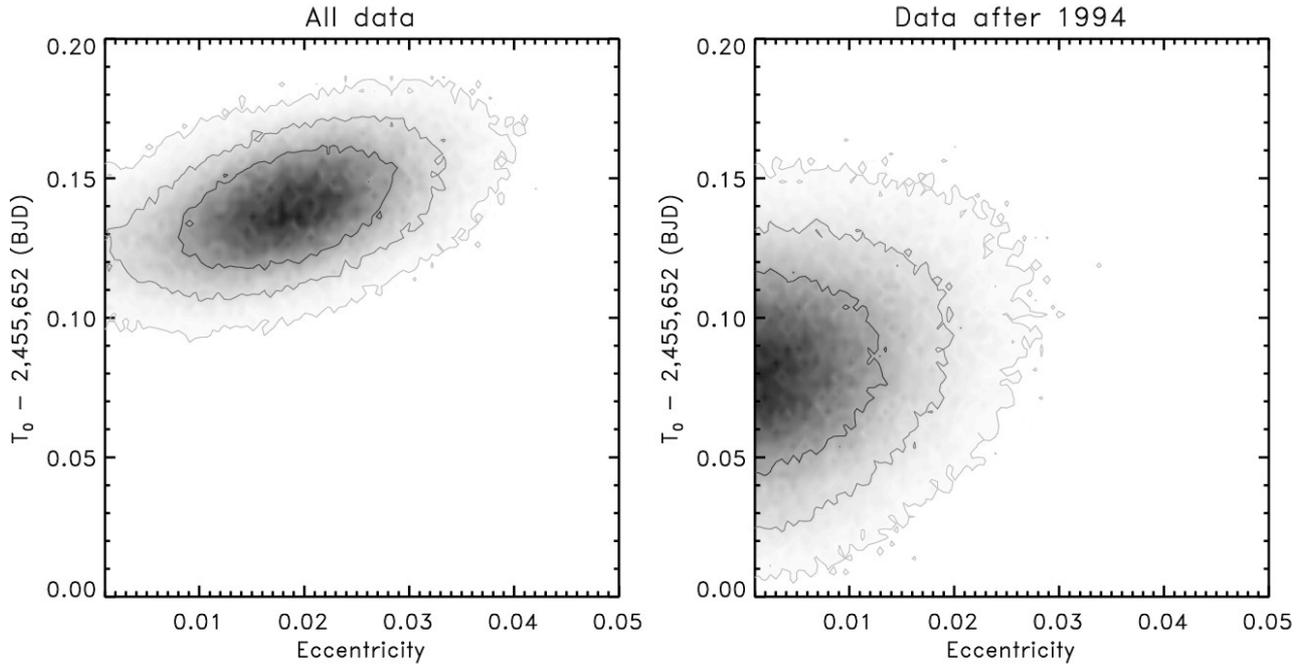

*Fig. S9: The MCMC density distribution of eccentricity (e) and time of inferior conjunction ($T_0$), as obtained from our independent fit of the available stellar radial velocity data of τ Boötis, with the left panel showing the result when all the measurements are included, and in the right panel only measurements taken after 1994. In the latter case, there is no evidence for a non-circular orbit. The contour levels correspond to the 1σ, 2σ and 3σ uncertainties in the MCMC distribution.*

## SI-5. The planet atmosphere

### SI-5.1 – Atmospheric models

We computed model spectra for the atmosphere of τ Boötis b, assuming a mean vertical profile in hydrostatic equilibrium. The profile is defined by two points in the temperature/pressure (T/P) space, ($T_1$, $p_1$) and ($T_2$, $p_2$), for which an isothermal atmosphere is assumed at $p > p_1$ and $p < p_2$, with temperatures $T_1$ and $T_2$ respectively. In between these two pressure levels, a constant temperature gradient is assumed as function of the logarithm of the pressure, $dT/d(\log p)$, except when an adiabatic lapse rate is assumed. In this last case, the temperature gradient is set to the adiabatic value in each one of the atmospheric layers between $p_1$ and $p_2$. A large range in Volume Mixing Ratios (VMR) for carbon monoxide, water vapour and methane is considered. The observed wavelength region is modelled including 2,225 CO lines, 11,631 $H_2O$ lines and 15,146 $CH_4$ lines. Data for CO and $H_2O$ are obtained from the HITEMP 2010 database[29], while that for $CH_4$ come from HITRAN 2008[24]. Although at high temperatures these catalogues could miss some of the weakest lines contributing to the continuum (especially the HITRAN database), all the strong lines which determine the high resolution spectrum are present. Gas opacities are calculated line-by-line using a Voigt profile. In addition, $H_2$-$H_2$ Collision-Induced Absorption (CIA) is also included. Note that the assumed temperature profiles, thought not realistic over the whole extent of a planetary atmosphere, can be effectively used to investigate the parameter space at the pressure levels probed by the molecular absorption lines, and are also easily reproducible by other groups that wish to perform similar analyses.

**SI-5.2 – Comparison of model spectra to the data**

When we compare the model spectra to the data, two model-dependent observables constrain the CO VMR and the atmospheric T/P profile: 1) the total strength of the correlated signal, and 2) the relative strength of the signals between arrays #2 and #3.

The first observable, the total strength of the correlated signal, is not directly related to the true strength of the molecular lines. This is because the data analysis treats each pixel in the spectrum differently, since each column in the data matrix is normalised by its standard deviation as part of the analysis (see SI-2.2). Therefore, for comparing the observed signal strength with that expected for a certain model atmosphere, we inject the reverse model at the observed planet velocity in the data at the beginning of the data-analysis chain, such that the model spectrum is treated in the same way as the data. If the CO lines in the model spectrum are too weak, the observed signal will only be partially removed. If the lines are too strong, they will overcompensate the data, resulting in a positive (emission) cross-correlation signal. In order to scale a particular model to the spectrum, we need to make assumptions about the planet/star radii ratio (which is unknown because τ Boötis b is a non-transiting planet), and about the spectrum of the star. We adopt a value of $R_P = 1.15\ R_{Jup}$ for the planet radius, determined from the mean of the radii of all known transiting planets[21] within the mass range 3-9 $M_{Jup}$. For the stellar spectrum we use the NextGen model[30] with $T_{eff}$ = 6400 K, log($g$) = 4.5 (CGS units) and [Fe/H] = 0.5, assuming a stellar radius[15] of $R_S = (1.46 \pm 0.05)\ R_{Sun}$. Note that uncertainties in the planetary radius, also estimated from the previous sample of transiting exoplanets, affect the absolute level of the signal by ~35%, and consequently do not change the order of magnitude of our estimates.

For the second observable, the relative strengths of the signals in array #2 and #3 are compared to those expected for a model spectrum. This comparison is incorporated in the signal retrieval described in Section SI-2.3, since arrays #2 and #3 are summed with a weight determined from the template model. Therefore, the significance of the detection is already a measure of the second observable.